\documentclass[aps,prl,twocolumn,groupedaddress]{revtex4-1}
\usepackage{amsmath,amssymb,amsthm,verbatim}

\newtheorem{theorem}{Theorem}

\newcommand{\ket}[1]{|#1\rangle}

\newcommand{\op}[2]{|#1\rangle \langle #2|}

\newcommand{\LU}{\overset{\underset{\mathrm{LU}}{}}{\rightarrow}}

\begin{document}


\title{{\bf Deciding Unitary Equivalence Between Matrix Polynomials and Sets of Bipartite Quantum States}}


\author{Eric Chitambar$^1$}
\email[]{e.chitambar@utoronto.ca}
\author{Carl A. Miller$^2$}
\email[]{carlmi@umich.edu}
\author{Yaoyun Shi$^2$}
\email[]{shiyy@umich.edu}
\affiliation{$^1$Center for Quantum Information and Quantum Control (CQIQC), \\
Department of Physics, University of Toronto\\ Toronto, Ontario M5S 3G4, Canada\\
$^2$Department of Electrical Engineering and Computer Science\\
University of Michigan\\ 
2260 Hayward Street, Ann Arbor, MI 48109-2121, USA
}

\date{\today}

\begin{abstract}
In this brief report, we consider the equivalence between two sets of $m+1$ bipartite quantum states under local unitary transformations.  For pure states, this problem corresponds to the matrix algebra question of whether two degree $m$ matrix polynomials are unitarily equivalent; i.e. $UA_iV^\dagger=B_i$ for $0\leq i\leq m$ where $U$ and $V$ are unitary and $(A_i, B_i)$ are arbitrary pairs of rectangular matrices.  We present a randomized polynomial-time algorithm that solves this problem with an arbitrarily high success probability and outputs transforming matrices $U$ and $V$.  
\end{abstract}


\maketitle

\section*{Introduction}

With entanglement being one key component in the design and operation of quantum computers, it has become natural to treat entanglement as a resource which we extract from quantum systems and put to use.  Under this interpretation, much research has been devoted to quantifying the amount of entanglement present in the state of a given system \cite{Horodecki-2009a}.  However, it was soon realized that no single quantification or entanglement measure can fully capture a state's non-classical properties, and thus one must first stipulate a relative measure when asking how much entanglement some state possesses \cite{Horodecki-1998a}.  A common property of all meaningful measures is that entanglement between two subsystems cannot increase on average when manipulations are \textit{local}, or applied to each subsystem distinctly; \textit{global} actions are required to increase entanglement \cite{Vidal-2000a}.  Because of the reversibility in unitary evolution, an immediate consequence of this is that for all entanglement measures, entanglement remains constant under local unitary operations (LU).  As a result, studying LU equivalence is important since it identifies states that have the same amount of entanglement.  

With this motivation, we investigate the question of when two sets of bipartite states are simultaneously related by a local unitary operation.  More precisely, given two sets of states $\{\rho_0,...,\rho_m\}$ and $\{\sigma_0,...,\sigma_m\}$ shared between parties Alice and Bob, when is it possible for the duo to apply a fixed local unitary operation that pairwise transforms $\rho_i\LU\sigma_i$ for $0\leq i\leq m$?  Physically, this question arises when Alice and Bob are secretly given one of $m+1$ possible initial states, and they wish to know whether, with certainty, it is possible to obtain one of $m+1$ particular target states through local unitary operations alone.

In the specific case of just a single \textit{pure} state pair $\rho$ and $\sigma$, LU equivalence is decided by an equivalence in singular values.  Recently, significant progress has been made in the study of single-copy equivalence between multipartite pure states \cite{Kraus-2010a, Sawicki-2010a}.  Something not provided in these works, however, is a rigorous account of computational costs required to implement the described algorithms.  In the case of generic bipartite \textit{mixed} states, equivalence between $\rho$ and $\sigma$ is determined by a set of trace invariants \cite{Albeverio-2003a, Sun-2006a}, while the full solution to bipartite mixed state LU equivalence still remains open.  The generalization of these questions to simultaneous LU equivalence between multiple pairs of states has yet to be addressed, and such an investigation nicely complements previous work on simultaneous state transformations under global operations \cite{Chefles-1998a, Chefles-2000a, Feng-2005a} and simultaneous \textit{stochastic} local state transformations between two pairs of pure states \cite{Ji-2005a}.

On its own, the above question generalizes the purely linear algebraic problem of deciding for $m+1$ pairs of $d_1\times d_2$ matrices $(X_i, Y_i)$ whether there exists unitary matrices $U$ and $V$ such that $U X_i V^\dagger=Y_i$ for all $i$.  To our knowledge, this problem has not yet been studied either in the linear algebra community, although Radjavi has solved the special case of square matrices and $U=V$ \cite{Radjavi-1968a}.  The problem can be phrased in a manner better suited for deeper analysis by introducing degree $m$ matrix polynomials $\mathcal{P}(\lambda)=\sum_{i=0}^m\lambda^i X_i$ and $\mathcal{Q}(\lambda)=\sum_{i=0}^m\lambda^i Y_i$.  Two matrix polynomials are called unitarily equivalent if $U\mathcal{P} V^\dagger=\mathcal{Q}$, and we see that $U X_i V^\dagger=Y_i$ for all $i$ if and only if their corresponding matrix polynomials are unitarily equivalent.  We also note a more general notion of matrix polynomial equivalence in which $\mathcal{P}\sim\mathcal{Q}$ if there exists invertible constant matrices $A$ and $B$ such that $A\mathcal{P}B^{-1}=\mathcal{Q}$. 

In this report, we present a randomized polynomial-time algorithm that decides whether two sets of bipartite pure states can be made equivalent by a fixed local unitary operation.  For sets of $N$-partite mixed states, the algorithm can be used to decide whether each pair is simultaneous equivalent under the same \textit{unilocal} unitary operation.  These are special operations in which just a single party applies a local unitary while the other subsystems are left unperturbed.  Our algorithm applies to sets of any size and the probability of failure can be made arbitrarily small since the randomness arises from a polynomial identity testing subroutine in the algorithm.  The underlying technique of the algorithm also works to decide when two degree $m$ matrix polynomials are equivalent in the more general sense of invertible transforming matrices $A$ and $B$.

Finally, we note that our result will also decide general (not just unilocal) LU equivalence of generic bipartite mixed states, although the set of ``generic states'' in our case is different than those in \cite{Albeverio-2003a, Sun-2006a}.  Generic here means that the set of states to which our algorithm \textit{does not} apply has measure zero.  Specifically, our algorithm can be implemented on states that have distinct eigenvalues.  If $\rho=\sum_ic_i\op{\phi_i}{\phi_i}$ with $c_i > c_{i+1}$ and $\sigma=\sum_ic'_i\op{\phi_i'}{\phi_i'}$ with $c'_i > c_{i+1}'$, then $\rho$ and $\sigma$ are LU iff $c_i=c'_i$ and $\ket{\phi_i},\ket{\phi_i'}$ are LU equivalent for all $i$.  

The actual problem we will consider is a bit more general than the one described in the previous paragraphs and will be called the \textbf{Unitary Equivalence Problem} (UEP):
\begin{itemize}
\item[{}]  Suppose $G_1$ and $G_2$ are sub-$\mathbb{C}$-algebras of the rings $\mathbb{C}^{d_1 \times d_1}$ and $\mathbb{C}^{d_2 \times d_2}$, respectively.  For two sets of matrices $\{X_i\}_{i=0,...,m}$ and $\{Y_i\}_{i=0,...,m}$ with $X_i,Y_i\in\mathbb{C}^{d_1\times d_2}$, decide if there exists a unitary solution $U$ and $V$ to the system of equations
\begin{align}
\chi=\{&UX_iV^\dagger=Y_i:U\in G_1, V\in G_2\}.
\end{align}
\end{itemize}

The UEP formulation generalizes many different unitary equivalence problems.  For instance, if we let $G_1=\mathbb{C}^{d_1\times d_1}$ and $G_2=\mathbb{C}^{d_2\times d_2}$, we recover the question of whether there exists general unitaries $U$ and $V$ such that $UX_iV^\dagger=Y_i$ for all pairs $(X_i, Y_i)$.  If we furthermore consider $d_1=d_2$ with one pair of matrices both being the identity matrix $(I_{d_1},I_{d_1})$, the question becomes whether $UX_iU^\dagger=Y_i$ for all $i$.  An example of a nontrivial algebra $G_1$ is the set $\{M\otimes I_b: M\in\mathbb{C}^{a\times a}\}$ where $ab=d_1$. 

It is easy to see the connection between UEP and the simultaneous LU equivalence between bipartite states.  The states of a $d_1\times d_2$-dimensional bipartite system can be represented as vectors $\ket{\psi}$ in the product space $\mathbb{C}^{d_1}\otimes \mathbb{C}^{d_2}$, and linear operators on this space correspond to physical actions on the system.  By choosing some basis $\ket{i}_1$ and $\ket{i}_2$ for spaces $\mathbb{C}^{d_1}$ and $\mathbb{C}^{d_2}$ respectively, any state can be written as $\ket{\psi}=(I \otimes \psi)\ket{\Phi}$ where $\ket{\Phi}= \sum_{i=1}^d\ket{i}_1\ket{i}_2$.  This allows for a bipartite pure state $\ket{\psi}$ to be identified with the matrix $\psi\in\mathbb{C}^{d_1\times d_2}$ so that the transformation $\ket{\psi}\to(A\otimes B)\ket{\psi}$ corresponds to $\psi\to A\psi B^T$.  Consequently, simultaneous LU equivalence between states $\{\ket{\psi_i}\}_{i=0...m}$ and $\{\ket{\phi_i}\}_{i=0...m}$ amounts to whether $U\psi_i V^\dagger=\phi_i$ for all $i$.  For bipartite mixed states, the UEP is encountered only in the restricted setting of unilocal equivalence.  Since mixed states themselves are represented by elements in $\mathbb{C}^{d_1d_2\times d_1d_2}$, unilocal unitary equivalence between states $\rho$ and $\sigma$ is the question of whether $(U\otimes I_{d_2})\rho(U^\dagger\otimes I_{d_2})=\sigma$, which as noted above is an UEP instance.  Note that in the case of simultaneous unilocal equivalence of mixed states, the reduction to UEP applies to systems with an arbitrary number of parties.

\section*{The Algorithm}
As we will see in greater detail, the UEP can be solved by determining whether or not a particular system of quadratic equations has a nontrivial solution.  One strategy sometimes helpful for dealing with quadratic constraints is to relax the problem into a system of linear equations such that a solution to the new equations will solve the original with high probability.  We demonstrate this idea on the problem of deciding whether two $d_1\times d_2$ (assume $d_2\geq d_1$) matrix polynomials $\mathcal{P}=\sum_{i=0}^m\lambda^i X_i$ and $\mathcal{Q}=\sum_{i=0}^m\lambda^i Y_i$ are generally equivalent, i.e. $\mathcal{P}\sim\mathcal{Q}$.  In other words, does the system of equations 
\begin{align}
\chi_1=\{&AX_iB^{-1}=Y_i:A\in\mathbb{C}^{d_1\times d_1},\notag\\ 
&B\in\mathbb{C}^{d_2\times d_2}\;, 0\leq i< m\}
\end{align} have a nonzero solution for invertible $A$ and $B$?  Clearly $\chi_1$ has such a solution iff there are nonzero invertible solutions to 
\begin{align}
\chi_1'=\{&AX_i=Y_iB:A\in\mathbb{C}^{d_1\times d_1}, \notag\\
&B\in\mathbb{C}^{d_2\times d_2}\;, 0\leq i< m\}.
\end{align}
There are $O(md_2^2)$ linear equations in $\chi_1'$ which can be solved thus placing constraints on the $O(d_2^2)$ free variables of $A$ and $B$.  A matrix solution space to $\chi'_1$ is then generated by expressing $A\oplus B$ in terms of the remaining free variables, and $\chi_1$ has a solution iff there exists a nonsingular element in this space.  

A standard randomized algorithm for deciding whether a matrix subspace has a full rank element consists of evaluating the degree $O(d_2^2)$ real polynomial $|Det(A\oplus B)|^2$ for randomly selected values of the free variables.  The Schwartz-Zippel Lemma states that
for some $n$-variate polynomial $f(x_1,\cdots,x_n)$ over a field $\mathbb{K}$
and having degree no greater than $d$, if $f$ is not identically
zero, then $\text{Prob}[f(x'_1,\cdots,x'_n)=0]\leq\frac{d}{|X|}$
where each $x'_i$ is independently sampled from some finite set
$X\subset \mathbb{K}$ \cite{DeMillo-1978a,Schwartz-1980a,Zippel-1979a}. 

To use the Schwartz-Zippel Lemma for testing whether $|Det(A\oplus B)|^2$ is identically zero with success probability at least
$1-\frac{2d_2^2}{|X|}$, one
evaluates it on values randomly chosen from set $X\subset\mathbb{R}$ and decides a zero
identity if and only if the evaluation output is zero.  As any polynomial number of linear equations can be solved in a polynomial amount of time in order to obtain the space $A\oplus B$, we thus have an efficient method for deciding whether $\mathcal{P}\sim\mathcal{Q}$ up to any probabilistic degree of certainty.  We note that the Schwartz-Zippel technique can also be used in the study of bipartite entanglement distillation from a multipartite-party state \cite{Chitambar-2009a}.

To solve $\chi$, we work analogously to $\chi_1$ but with additional constraints enforced.  Consider the system 
\begin{align}
\chi'=\{&AX_i=Y_iB,X_iB^\dagger=A^\dagger Y_i :A,A^\dagger\in G_1,\notag\\
& B,B^\dagger\in G_2;, 0\leq i< m\}.
\end{align}
Then we have
\begin{theorem}
\label{thm:LU invertible system}
$\chi$ has a solution iff $\chi'$ has an invertible solution $A$ and $B$.
\end{theorem}
\begin{proof}
If such a solution for $\chi'$ exists, then $A^\dagger A X_i= X_i B^\dagger B$ and $AA^\dagger Y_i=Y_iBB^\dagger$.  But these equations imply $p(A^\dagger A) X_i= X_i p( B^\dagger B)$ and $p(AA^\dagger )Y_i=Y_ip(BB^\dagger)$ where $p$ is any polynomial function.  Let $x_i$ denote the distinct eigenvalues from the combined spectrums $\lambda(A^\dagger A)\cup\lambda(B^\dagger B)$.  Let $X$ be the Vandermonde matrix of the $x_i$, and $v$ the column matrix whose entries are $\sqrt{x_i}^{-1}$.  Then the entries of $X^{-1}v$ provide the coefficients of a polynomial $p(t)$ such that $p(A^\dagger A)=\sqrt{A^\dagger A}^{-1}$ and $p(B^\dagger B)=\sqrt{B^\dagger B}^{-1}$.  Note also that $p(A^\dagger A)\in G_1$ and $p(B^\dagger B)\in G_2$.  Define unitary matrices $U=A\sqrt{A^\dagger A}^{-1}\in G_1$ and $V=B\sqrt{B^\dagger B}^{-1}\in G_2$.  Then $UX_i=AX_i\sqrt{B^\dagger B}^{-1}=Y_i B\sqrt{B^\dagger B}^{-1}=Y_i V$.
\end{proof}

Matrix bases for $G_1$ and $G_2$ will contain no more than $d_2^2$ elements so that $\chi'$ represents $O(md_2^2)$ linear constraints on $O(d_2^2)$ free variables.  Indeed, two additional variable matrices $M$,$N$ can be introduced to $\chi'$ giving the equations $AX_i=Y_iB$, $X_iN=MY_i$, $B^\dagger=N$, $A^\dagger=M$, $A,M\in G_1$, and $B,N\in G_2$.  A solution matrix space $A\oplus B$ is generated, and like before, a polynomial identity test can be applied to decide with arbitrarily high probability whether this space contains a nonsingular element.  If a nonsingular element is found, use the $A$ and $B$ to form unitaries $U$ and $V$ as in Theorem \eqref{thm:LU invertible system}.

\section*{Conclusion}
In this article we have studied the general problem of determining when a set of matrix transformations can be simultaneously achieved by a left and right unitary action.  Physically, this corresponds to performing multiple transformations between bipartite pure states with the same local action so that the amount of entanglement remains unchanged.  Our analysis also extends to the situation of simultaneous unilocal unitary transformations on $N$-partite mixed states.  We have developed a polynomial-time randomized algorithm that decides the problem with high probability and also provides a unitary solution if it exists.  Related open questions concern simultaneous general LU equivalence between sets of bipartite mixed states and of states having more than three parties.  However, for this latter question, it appears the matrix polynomial and randomized techniques used above apply only to the types of LU equivalence considered in this report.

\begin{acknowledgments}
This work was supported in part by the National Science
Foundation of the United States under Awards~0347078, 0622033, 0502170, and 1017335.  E.C. is partially supported by CIFAR, CRC, NSERC, and QuantumWorks.
\end{acknowledgments}

\bibliography{EricQuantumBib}

\end{document}